\RequirePackage[hyphens]{url}
\documentclass[aps,prb,twocolumn,showkeys,superscriptaddress]{revtex4}
\usepackage[utf8]{inputenc}
\usepackage{amsmath}
\usepackage{hyperref}
\usepackage{graphicx}
\usepackage{appendix}
\usepackage{color}

\usepackage{etoolbox}
\makeatletter
\patchcmd{\NAT@citesuper}%
  {\textsuperscript{#1}}{\textsuperscript{[#1]}}{}{}
\makeatother

\begin{document}

\title{Numeric optimization for configurable, parallel, error-robust entangling gates in large ion registers}
\author{Christopher D. B. Bentley}
\author{Harrison Ball}
\author{Michael J. Biercuk}
\altaffiliation[Also ]{ARC Centre for Engineered Quantum Systems, The University of Sydney, NSW Australia}
\author{Andre R. R. Carvalho}
\author{Michael R. Hush}
\author{Harry J. Slatyer}
\affiliation{%
 Q-CTRL, 
 Sydney, NSW Australia \& Los Angeles, CA USA
}%
\date{\today}

\begin{abstract}
    We study a class of entangling gates for trapped atomic ions and demonstrate the use of numeric optimization techniques to create a wide range of fast, error-robust gate constructions.  Our approach introduces a framework for numeric optimization using individually addressed, amplitude and phase modulated controls targeting maximally and partially entangling operations on ion pairs, complete multi-ion registers, multi-ion subsets of large registers, and parallel operations within a single register.  Our calculations and simulations demonstrate that the inclusion of modulation of the difference phase for the bichromatic drive used in the M\o lmer-S\o rensen gate permits approximately time-optimal control across a range of gate configurations, and when suitably combined with analytic constraints can also provide robustness against key experimental sources of error.  We further demonstrate the impact of experimental constraints such as bounds on coupling rates or modulation band-limits on achievable performance.  Using a custom optimization engine based on TensorFlow we also demonstrate time-to-solution for optimizations on ion registers up to 20 ions of order tens of minutes using a local-instance laptop, allowing computational access to system-scales relevant to near-term trapped-ion devices.
\end{abstract}


\maketitle

\section{Introduction}

Quantum computing requires a universal set of high-fidelity gates that are fast, robust and scalable.\cite{DiVi00FP}
In trapped ion systems,\cite{CZ95PRL, Haff08PR} entangling gates are mediated by shared motional modes that are coupled to the qubit states through atom-light interactions. 
High-fidelity entangling gates many orders of magnitude faster than decoherence timescales have been demonstrated,\cite{Gaeb16PRL, Ball16PRL} and research continues on faster gate times while maintaining high fidelities.\cite{GZC03PRL, Bent15NJP, Scha18N} The dynamics for M\o lmer-S\o rensen-type operations is derived for the regime where gate timescales are slow compared to the trapping period, however errors arising from approaching this timescale can be mitigated by careful control~\cite{ShapiraarX2019}. Two-qubit M\o lmer-S\o rensen gates\cite{MolmerPRL1999,SorensenPRL1999,SorensenPRA2000} have been implemented with infidelity on the order of $10^{-3}$.\cite{Gaeb16PRL}  

Recently, a range of control protocols has been introduced to expand the functionality of these gates via modulation of the laser fields mediating the spin-motional interaction.  For instance, MS gates have demonstrated tremendous flexibility, permitting parallel couplings within large registers~\cite{Figg19N, GrzesiakarX2019} and using overlapping pairs~\cite{Lu19N, ShapiraarX2019, GrzesiakarX2019} via control modulation.  Moreover, the addition of control permits the introduction of noise and drift-robustness even in complex multi-ion settings.\cite{Hayes:2012, GreenPRL2015, Leun18PRL, Webb_2018,Zarantonello_2019, MilnePRApp2020}

The general theory for the controlled dynamics of M\o lmer-S\o rensen-type operations is well established (see e.g.~[\onlinecite{Lu19N}]), however special cases are typically explored in theory and experiment to simplify implementation or to make dynamical control more tractable. The most common dynamical control method employs modulation of the amplitude (with fixed phase) of the control drives.\cite{Zhu06EL, RoosNJP2008, Choi14PRL, Figg19N, GrzesiakarX2019, Zarantonello_2019}
This restriction to a real drive permits deeper analytical treatment of the gate conditions,\cite{Blum19arX} and reduces the degrees of freedom required for numerical optimization.  Accordingly, amplitude-modulated gates have been successfully implemented in a range of experiments including the execution of five parallel pairwise interactions within an 11-ion chain,\cite{GrzesiakarX2019} or in achieving a many-body 12-of-12 qubit gate; \cite{ShapiraarX2019} executing parallel gates of this nature is essential for algorithmic scalability when employing large or even mesoscale ion registers.  On the other hand, complex drives have been demonstrated experimentally using phase-modulation~\cite{Lu19N, MilnePRApp2020} or laser-detuning modulation.\cite{Leun18PRL}  However these have generally been limited to demonstrations with smaller registers, in part due to the challenge of efficient gate construction within a large control space.

In this paper, we demonstrate computational access to a general control framework leveraging modulation of complex control drives, and apply this framework to efficiently achieve a range of optimized error-robust gates in large ion registers.  Framing the gate conditions in this way allows numeric optimization of amplitude- and phase-modulated controls optimized for each individually-addressed ion within a register. This accommodates many-qubit and parallel operations within a single register, where the target relative phases $\psi_{jk}$ between each pair of ions $j$ and $k$ can be freely specified while ensuring qubit-motional decoupling.  We first present the theoretical framework for M\o lmer-S\o rensen-type operations, including introduction of an operational fidelity measure and error-robustness conditions. Next, we pose the numeric optimization problem subject to a variety of hardware-motivated constraints, before presenting results derived from a custom TensorFlow-based optimization package.\cite{Ball2020}
We demonstrate a range of high-fidelity, error-robust, and scalable control solutions for parallel many-body operations within ion-chains up to 20 ions. Comparative analysis reveals that the specific inclusion of a complex drive provides access to otherwise unachievable entangling-gate fidelities and reduced gate-durations for a broad range of laser detunings.  Finally, we study the scaling of computational resources and control parameters required to obtain solutions for different-length ion chains.

\begin{figure}[t!]
    \centering
    \includegraphics[width=\columnwidth]{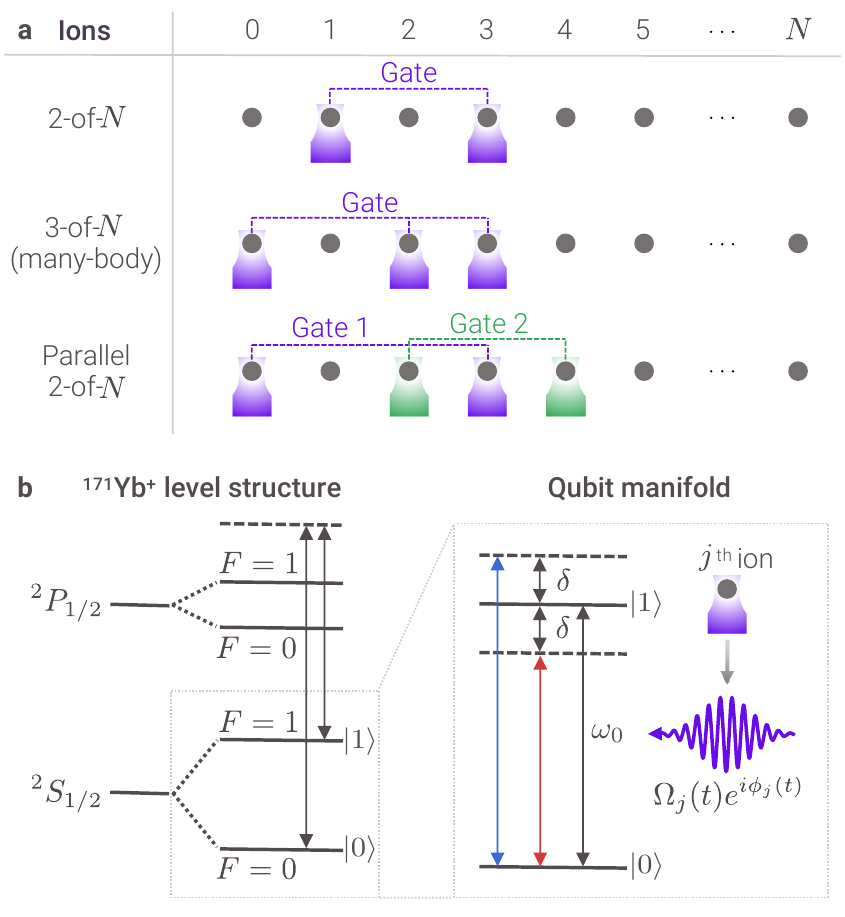} 
    \caption{Schematic depiction of gate configurability and modulation technique. (a) Different accessible gate configurations including pairs within large chains (top), many-body ($N>2$) arbitrary entanglement within chains (middle), and parallel gates within a chain (bottom). For simplicity we depict only two parallel pairwise gates, but arbitrary configurations within the chain are possible as shown {\em e.g.} in Figure~\ref{fig:largearray}. (b) Representative schematic level structure for Raman transitions, highlighting the laser beams which provide accessible controls for each individual ion.  In general, these individual modulated beams are combined with a single global laser to provide the beat-note spanning the qubit transition in a M\o lmer-S\o rensen-type gate. 
    \label{fig:concept}}
\end{figure}

\section{Operation dynamics and measures}

M\o lmer-S\o rensen-type multi-qubit operations employ bichromatic lasers which produce beatnotes detuned above and below the qubit transition frequency in order to achieve a generalized pairwise coupling, as shown in Figure~\ref{fig:concept}b for the specific example case of $^{171}$Yb$^{+}$ ions.  The laser detuning established by the Raman beatnote is kept close to the excitation frequencies of the motional modes used to couple the internal qubit states (Fig.~\ref{fig:concept}b). By M\o lmer-S\o rensen-\emph{type} operations we denote generalized forms of this operation that couple arbitrary pairs of qubits according to the unitary evolution operator: 
\begin{equation}
  U_c = e^{i \sum_{j=1}^N \sum_{k=1}^{j-1} \psi_{jk} \sigma_{x,j} \sigma_{x,k}}. \label{eq:target_unitary}
\end{equation} 
Here $j$ and $k$ are indices over ions, $\psi_{jk}$ is the target pairwise entangling phase and $\sigma_{x,j}$ is the Pauli $X$-operator acting on ion $j$. A maximally-entangling pairwise gate would have entangling phase $\psi_{jk} = \pi/4$ in this formulation. In the following we present a Hamiltonian-level description of the interactions and then frame the implementation subject to user-configurable constraints on the target operation.

\subsection{M\o lmer-S\o rensen dynamics}
We model the control problem for this gate, beginning with a conventional Hamiltonian description of the coupled dynamics of the internal and motional degrees of freedom for trapped ions, $H_0$:\cite{WinelandJRNIST1998}
\begin{equation}
    H_0 = \sum_{p=1}^M \hbar \nu_p \left( a_p^\dagger a_p + \frac{1}{2} \right) + \sum_{j=1}^N \frac{\hbar \omega_0}{2} \sigma_{z,j},
\end{equation}
where motional mode $p$ has frequency $\nu_p$, and the $N$ trapped ions have an internal qubit transition at frequency $\omega_0$.
We denote Pauli $k$-operators for ion $j$ as $\sigma_{k,j}$.

In the rotating frame with respect to $H_0$, the interaction Hamiltonian for M\o lmer-S\o rensen-type operations is given by:\cite{GreenPRL2015}
\begin{equation}
    H_I (t) = i \hbar \sum^N_{j=1} \sigma_{x,j} \sum^M_{p=1} \left( - \beta_j^{p*} (t) a_p + \beta_j^p (t) a_p^\dagger \right). \label{eq:ions_greenform}
\end{equation}
The term coupling ion $j$ to motional mode $p$ is given by:
\begin{equation}
    \beta_j^p (t) = \eta_j^p \frac{\gamma (t)}{2} e^{i \delta_p (t) t}, \label{eq:beta}
\end{equation}
where $\eta_j^p \equiv \eta_p b_j^{(p)}$ with $\eta_p$ the Lamb-Dicke parameter and ion-mode participation eigenvectors $b_j^{(p)}$.\cite{Jame98APB}
The relative detuning from the $p$th mode is $\delta_p (t) = \nu_p - \delta (t)$ with the laser frequency detuned by $\delta(t)$ from the qubit transition $\omega_0$. We represent the complex drive $\gamma (t) = \Omega (t) e^{i \phi (t)}$, with Rabi frequency $\Omega (t)$ and phase $\phi (t)$.

The interaction Hamiltonian is valid when several approximations hold.
First, it is necessary that phase-space (or equivalently ion) displacements remain small, such that:
\begin{equation}
    \langle (k x_j)^2 \rangle_{\rho_{\text{mot}}}^{\frac{1}{2}} \ll 1
    \quad \forall j, t,
\end{equation}
where $x_j$ is the displacement operator for ion $j$, $k$ is the addressing radiation wavevector, and $\rho_{\text{mot}}$ is the motional state of the ions.\cite{WinelandJRNIST1998}
Note that $k x_j = \sum_p \eta_j^p (a_p + a_p^\dagger)$.
Second, the protocol involves a pair (or pairs) of laser frequencies that we denote with subscripts $a$ and $b$: the laser pair has opposite detunings $\delta_a (t) = \delta (t)$, $\delta_b (t) = -\delta (t)$, and phases $\phi_a (t) = \phi (t)$, $\phi_b (t) = -\phi (t) + \pi$.
Finally, detunings $\delta (t)$ should be close to $\nu_p$, such that a rotating wave approximation eliminates carrier transitions.

The dynamical equations may be generalized to accommodate individual drives for different ions,\cite{Figg19N,GrzesiakarX2019,Lu19N} moving beyond the shared expression $\gamma(t)$ in equation~(\ref{eq:beta}). For ion-dependent complex drives, we transform $\gamma (t) \rightarrow \gamma_j (t)$ for the $j$th ion, with commensurate transformations $\phi (t) \rightarrow \phi_j (t)$ and $\Omega (t) \rightarrow \Omega_j (t)$.  

As highlighted in Fig~\ref{fig:concept}b, this ion-specific complex drive, induced by the Raman lasers, represents the key ``control knob'' in our possession.  It is this parameter over which we will perform optimization as outlined in section~\ref{sec:opt}. In this manuscript we fix the laser detuning in order to facilitate the numeric optimization described in section~\ref{sec:opt}, though in principle this parameter may be transformed in the same way. 

The unitary operator resulting from equation~(\ref{eq:ions_greenform}) can be written as time-ordered infinitesimal (state-dependent) displacement operators, from which (up to global-phase terms) we obtain:
\begin{align}
    U (\tau) &= \exp \left( \sum_{j=1}^N \sigma_{x,j} B_j (\tau)  \nonumber \right. \\
    & \left. \quad + \quad i \sum_{j=1}^N \sum_{k=1}^{j-1} \left( \phi_{jk} (\tau) + \phi_{kj} (\tau) \right) \sigma_{x,j} \sigma_{x,k} \right), \label{eq:ions_unitary}
\end{align}
\begin{align}
    B_j (\tau) &\equiv \sum_{p=1}^M \left( \eta_j^p \alpha_j^p (\tau) a_p^\dagger - \eta_j^{p*} \alpha_j^{p*} (\tau) a_p \right),
\end{align}
\begin{align}
    \phi_{jk} (\tau) &\equiv \text{Im} \left[ \sum_{p=1}^M \int_0^\tau dt_1 \int_0^{t_1} dt_2 \beta_j^p (t_1) \beta_k^{p*} (t_2) \right]. \label{eq:phase}
\end{align}
Here, ion $j$'s contribution to the displacement of mode $p$ in phase space is given by
\begin{equation}
    \eta_j^p \alpha_j^p (t) = \eta_j^p \int_0^t dt' \frac{\gamma_j (t')}{2} e^{i \delta_p t'}. \label{eq:ions_trajdisplacement}
\end{equation}

\subsection{Target operations and fidelity metrics}
Our specific target is the achievement of high-fidelity operations under arbitrary couplings between ions within an $N$-ion register.  These couplings can be achieved for an individual pair, for multiple pairs in parallel, or as many-body ($M$-of-$N$) operations, respectively, as depicted in Figure~\ref{fig:concept}a.  This introduces two control targets in our problem.  First, we desire arbitrary and specifiable relative phases between ions $j$ and $k$. Referring to the target unitary in equation~(\ref{eq:target_unitary}), we thus require that the acquired phases satisfy
\begin{equation}
    \phi_{jk} (\tau) + \phi_{kj} (\tau) = \psi_{jk} \label{eq:ions_phaseconditions}
\end{equation}
\noindent for a gate of duration $\tau$. Next, we require elimination of qubit-motional entanglement at the completion of the operation. The residual qubit-motional entanglement is eliminated by ensuring that
\begin{equation}
    \alpha_j^p (\tau) = 0 \quad \forall j,p. \label{eq:ions_motionconditions}
\end{equation}

We quantify performance using the operational fidelity, which incorporates both entangling-phase and residual-motional-entanglement errors for a diverse range of operations:
\begin{equation}
    \mathcal{F}_{av} = \left| \frac{1}{D} \text{Tr} [E] \right|^2,
\end{equation}
with $E = U_c^\dagger U (\tau)$ and where $E$ has dimension $D$. 
Up to second order in the motional and phase error terms from equations~(\ref{eq:ions_motionconditions}) and~(\ref{eq:ions_phaseconditions}), $\alpha_j^p (\tau)$ and 
\begin{equation}
    \epsilon_{jk} \equiv \psi_{jk} - (\phi_{jk} (\tau) + \phi_{kj} (\tau)),
\end{equation}
respectively, we obtain:
\begin{align}
    \mathcal{F}_{av} &= \left| \left( \prod_{j=1}^N \prod_{k=1}^{j-1} \cos (\epsilon_{jk}) \right) \right. \nonumber \\
     & \left. \times 
    \left( 1 - \sum_{p=1}^M \sum_{j=1}^N \left[ \left|\eta_j^p \right|^2 \left|\alpha_j^p (\tau)\right|^2 \left(\overline{n}_p+\frac{1}{2} \right) \right] \right) \right|^2, \label{eq:fidelity}
\end{align}
as outlined in the Supporting Information (SI).
We have taken the expectation value of the ion motion with respect to a separable thermal product state with mean phonon occupation $\overline{n}_p$ in mode $p$.

In the following section, we introduce our optimization methodology and discuss our implementation of the robustness conditions.

\section{Obtaining control solutions: optimization methodology} \label{sec:opt}

\subsection{Optimization framework}
To obtain error-robust and high-fidelity control solutions, we consider temporal modulation of the complex drives $\gamma_j(t)$ for each ion $j$.
We choose a piecewise-constant basis to characterize the dynamics, defining the segmentation:
\begin{equation}
    \gamma_j (t) = \sum_{k=1}^S \chi_{A_k} (t) \gamma_{j,k} = \sum_{k=1}^S \chi_{A_k} (t) \Omega_{j,k} e^{i \phi_{j,k}},
\label{eq:piecewiseoptimization}
\end{equation}
\noindent where segment $k$ is defined over the interval $A_k = [t_k, t_{k+1}]$, and $\chi_{A_k}$ is the indicator function that takes a value of 1 for $t \in A_k$ and 0 otherwise.  The operation begins at $t_1 = 0$ and finishes at $t_{S+1} = \tau$.
Note that one may follow the same procedure using different basis functions such that the control degrees of freedom are time-independent. 

Using this basis, we rewrite the entangling-phase-accumulation equation~(\ref{eq:phase}) and motional-displacement equation~(\ref{eq:ions_trajdisplacement}):
\begin{align}
    \phi_{mn}(\tau) &= \text{Im} \left[ \mathbf{u}_m^T P^{m,n} \mathbf{u}_n^* \right] \quad \forall m,n, \label{eq:phase_segs} \\
    \boldsymbol{\alpha}_n(\tau) &= M \mathbf{u}_n \quad \forall n, \label{eq:mot_segs}
\end{align}
where $n$ and $m$ are ion indices, $\mathbf{u}_n$ is the vector of controls such that element $k$ is the $k$th piecewise segment value $\gamma_{n,k}$, and entry $p$ of the vector $\boldsymbol{\alpha}_n$ is $\alpha_n^p$.
The matrices $M$ and $P^{m,n}$ have elements given by:
\begin{align}
     M_{p,k} &= \frac{1}{2} \int_{A_k} dt e^{i \delta_p t}, \\
     P^{m,n}_{k,l} &= \sum_{p=1}^M \frac{\eta_m^p \eta_n^p}{4} \int_{A_k} dt_1 e^{i \delta_p t_1} \int_{A_l \cap [0,t_1]} dt_2  e^{-i \delta_p t_2} ,
\end{align}
respectively. 
Here $p$ is an index over motional modes, and $k$ and $l$ are indices over segments.

To obtain control solutions, we apply a custom gradient-based optimization engine~\cite{Ball2020} built on TensorFlow to minimize the operation error. To this end we minimize the cost function $\mathcal{C}$ defined as:
\begin{equation}
    \mathcal{C} = \sum_{\substack{j=1 \\ k<j}}^N (\epsilon_{jk})^2 + \sum_{j=1}^N \sum_{p=1}^M |\alpha_j^p (\tau) |^2.
\end{equation}
The terms included, $(\epsilon_{jk})^2$ and $|\alpha_j^p (\tau) |^2$, are proportional to the lowest-order infidelity contributions, for each mode $p$ and ions $j, k$.
Minimizing this simpler functional form provides better performance than using the full functional form of the infidelity.
Using equations~(\ref{eq:phase_segs}) and~(\ref{eq:mot_segs}), we obtain quadratic and linear expressions for $\epsilon_{jk}$ and $\alpha_j^p (\tau)$ in terms of our control degrees of freedom, respectively.

Given our control and optimization framework, we may impose additional physical constraints on the free variables as part of the optimization problem.
This includes bounding the rate-of-change of the drive phase and amplitude (Figure~\ref{fig:robust} and corresponding to band limits in hardware), fixing the phase or amplitude (Figure~\ref{fig:domains}), sharing the same drive parameters between arbitrary ions in the chain (Figure~\ref{fig:domains}), or incorporating generic linear-time-invariant filters on control transmission.\cite{Ball2020,QCTRL_Bandlimited}

\subsection{Integration of error-robustness} \label{sec:robust}

We can analyze gate error-robustness, and reduce the error-susceptibility of optimized controls, by modelling the impact of common noise terms on the dynamic evolution of the system.  Here we focus on several different error processes that are commonly encountered in laboratory environments, ranging from trap instability and laser frequency drift to systematic timing errors.

Beginning with dephasing, this form of error can arise from imperfect calibration or drift in the motional mode frequencies $\nu_p$ as trapping potentials frequently vary in time.\cite{Hayes:2012, ShapiraarX2019}  Error in a given mode frequency $\nu_p \rightarrow \nu_p + \epsilon_p$ becomes a shift in the relative detuning $\delta_p \rightarrow \delta_p + \epsilon_p$, which impacts the mode closure:
\begin{equation}
    \bar{\alpha}_j^p(\tau) = \int_0^\tau dt \frac{\gamma_j(t)}{2} e^{i (\delta_p + \epsilon_p) t}. \label{eq:dephasing_alpha}
\end{equation}
In order to compensate the effect of quasi-static noise on mode trajectory closure to first order, we require: 
\begin{align}
    0=\left. \frac{d \bar{\alpha}_j^p(\tau)}{d \epsilon_p}\right|_{\epsilon_p = 0} &= i \int_0^\tau dt \frac{\gamma_j(t)}{2} e^{i \delta_p t} t, \label{eq:ions_dephasing_integral} \\ 
    &= i\tau \alpha_j^p(\tau) - i \int_0^\tau dt \alpha_j^p(t).
\end{align}
The term proportional to $\alpha_j^p(\tau)$ is set to zero in the usual motional conditions for an operation, and the integral over $\alpha_j^p(t)$ can be set to zero as an additional robustness condition, as in~[\onlinecite{ShapiraarX2019}]. 
Since $\alpha_j^p(t)$ is proportional to the displacement of ion $j$ in mode $p$ at time $t$, this condition is equivalent to setting the center of mass of ion $j$'s contribution to mode $p$'s phase space trajectory to zero.
When the center of mass is set to zero for each ion's contributions to phase space trajectories, the residual motion condition (trajectory closure) can be satisfied by enforcing symmetry in the controls as described in~[\onlinecite{MilnePRApp2020}].
This work found that robustness to both quasi-static and zero-mean fluctuating dephasing noise processes can be obtained by setting the center of mass of each motional mode's phase space trajectory to zero.

Dephasing noise can also arise from errors in the laser-pair detunings such that $\delta_a \neq - \delta_b$.
The gate dynamics can be rederived with this detuning asymmetry, as shown in the SI, where we find that the robustness conditions derived above also provide robustness to relative detuning noise.
  
We next consider systematic timing errors such that the control pulses are scaled by a uniform factor $(1 + \epsilon_t)$.
This affects the mode closure conditions in the following way:
\begin{equation}
    \bar{\alpha}_j^p (t_f=(1+\epsilon_t) \tau) = \int_0^{\tau (1+\epsilon_t)} dt \frac{\gamma_j(t/(1+\epsilon_t))}{2} e^{i \delta_p t},
\end{equation}
and transforming $t \rightarrow t' \equiv t/(1+\epsilon_t):$
\begin{equation}
    \bar{\alpha}_j^p (t'_f=\tau) = \int_0^{\tau} dt' (1+\epsilon_t) \frac{\gamma_j(t')}{2} e^{i \delta_p (1+\epsilon_t) t'},
\end{equation}
the mode closure impact is proportional to equation~(\ref{eq:dephasing_alpha}) with a dephasing shift $\epsilon_p = \epsilon_t \delta_p$.
This equivalence means that a control scheme satisfying the dephasing robustness conditions to a given order is also robust to timing errors to that same order.  

To apply error-robust optimization with respect to these noise sources, we require that the residual phase space displacements are zero as in equation~(\ref{eq:ions_motionconditions}), and that the integral (or center of mass) of each phase-space trajectory is zero.  The center of mass conditions can be written in a linear form with respect to the controls:
\begin{align}
    \mathbf{0} &= R \mathbf{u}_n \quad \forall n, \\
    R_{p,k} &= \int_0^T dt \int_{A_k \cap [0,t]} dt_1 e^{i \delta_p t_1}.
    \label{eq:com_conditions}
\end{align}
where $n$ is an index over ions, and the matrix elements $R_{p,k}$ of $R$ are defined in the second equation above.  Here $p$ is an index over motional modes, and $k$ is an index over segments.
If these conditions are satisfied, the closure of the phase space trajectories (satisfying the residual displacement conditions) can also be enforced by imposing symmetry in the drives across the temporal midpoint of the gate operation. \cite{MilnePRApp2020}    
For piecewise-constant drives with variable amplitude and phase, the symmetry can take the form:
\begin{align}
    \Delta \phi_{j,n+1} &= \Delta \phi_{j,S-(n+1)}, \nonumber \\
    \Omega_{j,n} &= \Omega_{j,S-n} \label{eq:ions_symmconditions}
\end{align}
where $\Delta \phi_{j,n} = (\phi_{j,n} - \phi_{j,n-1})$, $\Omega_{j,n}$ and $\phi_{j,n}$ are the fixed amplitude and phase for the $j$th ion over the $n$th drive segment, and $S$ is the number of segments in the drive. 
We note that the number of segments can be set independently for different ion-specific drives, as each drive modulation pattern is reflected individually to satisfy the symmetry conditions. We thus achieve error-robust solutions using a combination of symmetry and numerical optimization approaches.

We have derived and implemented robustness conditions particularly for laser-detuning or equivalent noise sources; one may alternatively consider laser amplitude fluctuations. Previous work~\cite{MilnePRApp2020} has demonstrated that in the ensemble average, zero-mean temporally fluctuating processes may be suppressed by the same prescription, but sensitivity to quasi-static errors in individual gates remains.  This is evident when considering the entangling phase equation~(\ref{eq:phase}) which shows that quasi-static errors of the form $\Omega(t) \rightarrow s\Omega(t)$ directly induce the acquired entangling-phase rescaling $\phi_{jk}(\tau) \rightarrow s^2 \phi_{jk}(\tau)$.  Such entangling-phase errors dominate infidelity contributions arising from residual motional entanglement and remain the subject of future work.

\begin{figure}[tbh]
    \centering
    \includegraphics[width=\columnwidth]{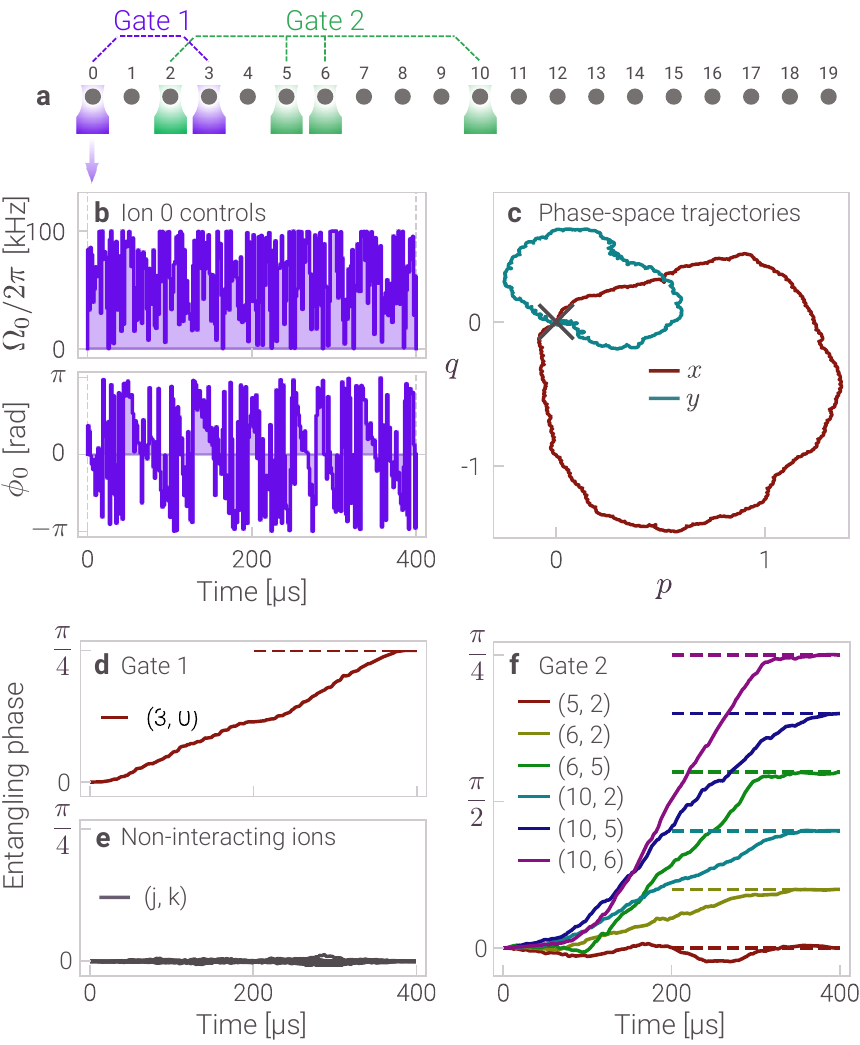}
    \caption{Simultaneous 2- and 4-qubit gates in a 20-ion chain, with infidelity $\mathcal{I}=1.8 \times 10^{-7}$.
    (a) Schematic of the operation: the ions are individually addressed with 256-segment drives. Gate 1 maximally entangles ions 0 and 3.
    Gate 2 produces pre-specified pairwise target phases between ions 2, 5, 6 and 10.
    (b) Optimized drive (modulated amplitude and phase) for ion 0. The maximum permitted Rabi rate for each drive is $\Omega_\text{max}/2\pi = 100$~kHz.
    (c) Phase space trajectories for the maximally-displaced motional modes on the $x$ and $y$ transverse trap axes.
    The final displacements at the end of the gate are marked with a cross.
    (d-f) Phase dynamics during the gate operations. Target phases for each ion pair are marked with dashed lines and color-coded to match the associated ion pair.
    Non-interacting ion pairs (j, k) (those not entangled by a target gate) have a target relative phase of 0.
    Note that additional simulation details are included in the Supporting Information (SI).
    \label{fig:largearray}}
\end{figure}

\section{Optimization results and performance analysis}

\begin{figure}[t!]
    \centering
    \includegraphics[width=\columnwidth]{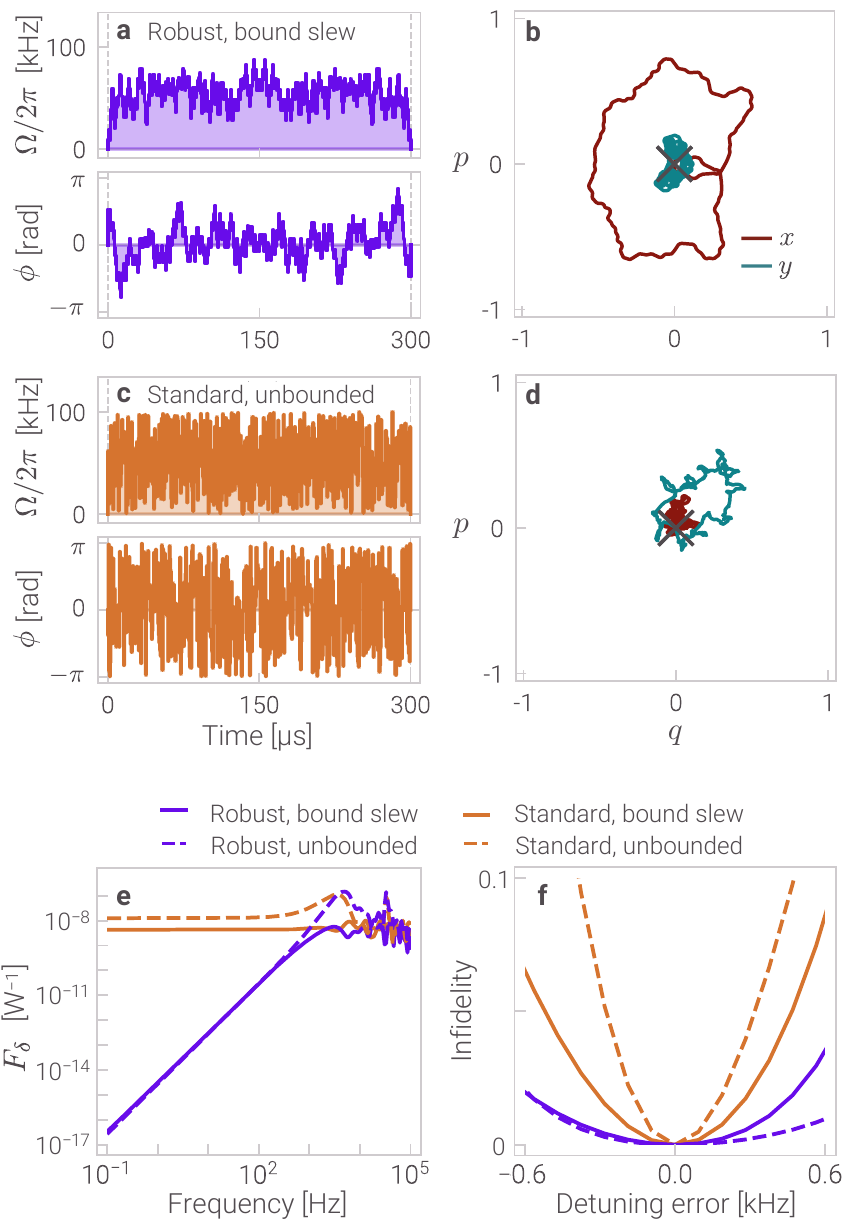}
    \caption{
    Controls, dynamics and robustness under different constraints for a 2-of-2 qubit maximally-entangling gate. 
    The same drive addresses both ions, with $\Omega_\text{max}/2\pi = 100$~kHz.
    (a) Robust drive with a bound rate of change for both the modulus and phase: $\Delta \Omega / 2 \pi \leq 10$~kHz, $\Delta \phi \leq \pi/8$ between segments with duration $\sim 0.6$~$\mu$s.
    The operational infidelity is $3.7 \times 10^{-9}$.
    (b) Phase space trajectories for the center-of-mass mode of each axis under the robust, physically-constrained drive in (a).
    A cross marks the endpoint of each trajectory at the origin.
    (c) Unconstrained optimized drive for the operation, with operational infidelity $1.5 \times 10^{-12}$.
    (d) Phase space trajectories for the center-of-mass mode of each axis under the unconstrained drive in (c).
    (e) Filter functions displaying susceptibility to dephasing noise, as described in the main text.
    (f) Quasi-static scans of dephasing noise for 'Robust' and 'Standard' (non-robust) optimized control solutions, with and without bound slew rates.
    \label{fig:robust}}
\end{figure}

\begin{figure}[t!]
    \centering
    \includegraphics[width=\columnwidth]{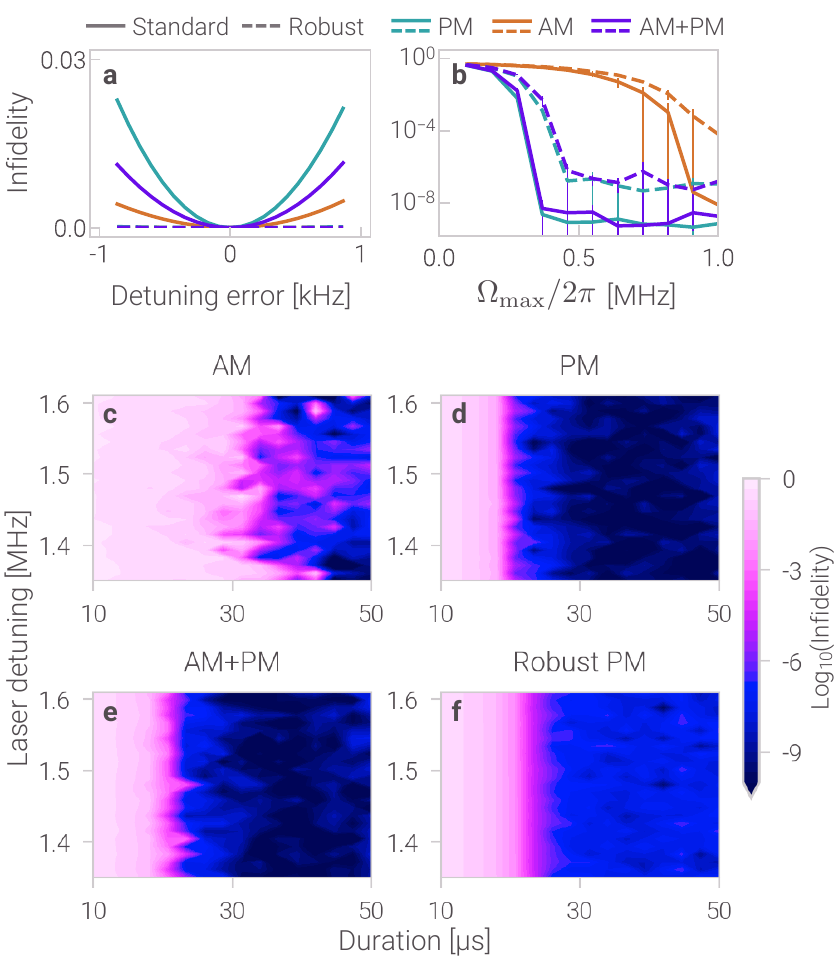} 
    \caption{Control solution analysis for a 2-of-5 qubit maximally-entangling gate with a shared (128-) 64-segment (robust) drive addressing ions 0 and 1.
    (a) Quasi-static scans of dephasing noise for instances of different control schemes: standard and robust amplitude-modulated (AM), phase-modulated (PM), amplitude- and phase-modulated (AM$+$PM) schemes. Robust gate infidelities overlap near zero for the displayed detunings.
    (b) Infidelity with maximum Rabi rate $\Omega_\text{max}$: this upper bound on $\Omega$ is applied to each control optimization.
    Error bars display a standard deviation from the mean (solid line) over ten runs of the optimizer (where each run involves five optimization instances).
    In (a,b) the detuning and gate duration are fixed at $\delta=1.365$~MHz and $\tau=50$~$\mu$s, respectively.
    (c-f) Optimized control infidelities for scans over laser detuning $\delta$ and gate time $\tau$, with $\Omega_\text{max}/2\pi = 1$~MHz. Different control configurations are displayed in each subfigure.
    \label{fig:domains}}
\end{figure}

An example optimization highlighting various capabilities of this framework is presented in Figure~\ref{fig:largearray}.
Here we optimize two parallel asymmetric gates on a chain of 20 ions, as shown in the schematic in Figure~\ref{fig:largearray}a, which achieve infidelity $\mathcal{I} = 1.8 \times 10^{-7}$.
The first gate ('Gate 1') is a maximally-entangling two-qubit gate between ions 0 and 3 in the chain (indexing from 0).
The second gate ('Gate 2') is a four-qubit gate on ions 2, 5, 6 and 10 that prepares user-defined relative phases between different sub-pairs in steps of $\pi/10$.
These choices of relative phases are configurable and were chosen arbitrarily to highlight the freedom inherent in the optimization.
Figure~\ref{fig:largearray}b displays the optimized drive for ion 0; the control for each ion varies rapidly between discretized time segments, exploiting the full parameter space afforded to the optimizer in achieving the target performance.  Controls for other ions are similar in overall appearance but vary in their detailed prescription.

The performance of the gate can be explored visually through simulation of both the phase-space motional dynamics (Figure~\ref{fig:largearray}c) and entangling phase for different ion pairs as a function of time during the gate (Figure~\ref{fig:largearray}d-f).  As expected, for two transverse motional modes illustrated here, both make a complex excursion in phase space before returning to the origin at the end of the gate, indicating efficient qubit-motional decoupling.  Similarly, we observe that the pairwise entangling dynamics for both Gate 1 and Gate 2 achieve target phases for each pair, and qubit pairs not involved in a gate have entangling phase restored to zero at the end of the operation.
These dynamics validate that the target relative phases and closed phase space trajectories are achieved by the end of the operation, despite the complexity of the simultaneously executed operations.

We may incorporate robust-performance constraints as well as physically motivated limitations on the form of the resultant controls through the optimization procedure.  We illustrate these capabilities for a 2-of-2 ion maximally-entangling gate with shared addressing in Figure~\ref{fig:robust}.  A key consideration in implementation is the response time of either RF signal generators or optical modulators employed in gate implementation, the experimental impact of which was treated in~[\onlinecite{MilnePRApp2020}].  In order to accommodate hardware constraints the optimization may include effective band limits implemented through a number of filtering techniques.

In Figure~\ref{fig:robust}a we illustrate one example hardware-compatible constraint based on limiting the time-derivative of the modulation profile, which we term a ``bound-slew-rate'' control.\cite{Ball2020} The bound-slew-rate control waveform achieves an infidelity $3.7\times 10^{-9}$, despite the substantial differences in allowable waveform relative to the unconstrained solution presented in Figure~\ref{fig:robust}c.
Phase space trajectories for the respective controls are displayed in Figure~\ref{fig:robust}b and~\ref{fig:robust}d, and reflect the limit on allowable modulation bandwidth through smoothing of the trajectories in Figure~\ref{fig:robust}b.  A variety of other smoothing filters could be considered, and are compatible with the optimization engine as described in~[\onlinecite{QCTRL_Bandlimited}], including arbitrary linear time-invariant filters which capture measured modulator responses, etc.

We demonstrate the error-robustness of these and two additional 2-of-2 gate optimizations using conventional analytic techniques in robust control.\cite{Ball2020}  First, for both the bound-slew-rate and unbounded optimizations we calculate the filter functions for gate variants designed to either simply enact the target gate or include robustness to detuning noise.  The filter function serves as a proxy measure for noise admittance as a function of noise frequency, and is experimentally validated for single-qubit gates~\cite{SoareNatPhys2014} and multi-qubit M\o lmer-S\o rensen gates.\cite{MilnePRApp2020}  A robust control will suppress noise at low frequencies, indicated by a filter function which takes small values in this range.  In Figure~\ref{fig:robust}e we observe that both bound-slew-rate and unbounded controls designed to be robust (purple lines) exhibit suppression of noise sensitivity at low frequencies.  By contrast the standard controls exhibit broadband noise susceptibility up to a frequency commensurate with the inverse gate time.

Similarly, we evaluate the robustness of the gates to quasi-static detuning errors (Figure~\ref{fig:robust}f) via calculation of gate infidelity in the presence of fixed detuning offsets.  Here we see that as a function of offsets from ``ideal'' laser settings (zero on the x-axis), gate infidelity will increase at varying rates depending on the specifics of gate construction.  The range of laser detuning over which infidelity remains low serves as an effective measure of error-robustness. The standard control solutions (orange) both exhibit a narrow range of detunings allowing high-fidelity implementation.  By contrast, the robust solutions exhibit a broad range of ``flat'' infidelity around zero detuning error, indicating that small drifts will not substantially degrade operational fidelity.  These results hold with or without bounds on the slew-rates for the controls.
The effective reduction of detuning-induced infidelity using our robust methodology is also displayed for 2-of-5 qubit gates in Figure~\ref{fig:domains}a, for different control schemes.

The demonstrations above have shown optimized controls utilizing complex drives, where both the amplitude and phase are modulated in time with the aim of achieving low gate infidelities. We now highlight the applicability of this methodology in achieving high-fidelity, short-time gates.

High-fidelity control solutions can be achieved for different gate time and laser detuning `domains' depending on the degrees of freedom in the control. As an example these domains are displayed in Figure~\ref{fig:domains}c-f for a maximally-entangling 2-of-5 qubit gate, using different modulation protocols:
amplitude-modulated (AM), phase-modulated (PM), amplitude- and phase-modulated (AM$+$PM) and robust phase-modulated controls (Robust PM).  Here, dark regions represent high-fidelity gate implementations that have been found by the optimizer while light regions show gate implementations exhibiting larger errors.  As expected, as gate durations decrease it becomes more challenging for the optimizer to find high-fidelity solutions, and below a certain threshold no high-fidelity gates may be achieved for a fixed maximum Rabi rate. 
In our calculations we find that both the high-fidelity domain and its boundary for AM controls routinely exhibit substantial structure yield an approximate minimum-gate-duration threshold nearly 50\% larger than gate constructions incorporating phase modulation. In the latter cases the optimal gate duration (for a given target infidelity) is reduced but also appears to depend only weakly on the choice of detuning.
It is interesting that AM$+$PM controls have a slightly reduced low-infidelity domain compared with the PM case despite being a super-set (any valid PM control is also a valid AM$+$PM control); this may simply be a manifestation of an underconstrained optimization problem exhibiting local minima.  Finally, we note that despite the reflection of controls (using twice as many segments) required to ensure robustness we observe only a marginal change in the threshold gate-duration before achieving high-fidelity gates when incorporating a robustness constraint.

Another practical consideration for gate implementation is the drive power requirement of a given scheme.
In Figure~\ref{fig:domains}b we display the achievable infidelity for a 2-of-5 qubit gate and different modulation schemes as a function of the permitted maximum drive power.
Again, we observe that the solution incorporating only AM is most restrictive; the optimized controls require higher drive power to reach infidelity below any given threshold.  The addition of phase modulation reduces required drive power by approximately $2\times$, whether used on its own or in combination with amplitude modulation.  In all cases we have considered, the addition of robustness constraints increases the maximum drive-power requirements ($\sim$10-20$\%$) and limits the best achievable infidelity due to the segment number in the controls.  In the presence of noise, however, the lower susceptibility of the robust solution can quickly outweigh this advantage of the ideal 'Standard' operations (as displayed in Figure~\ref{fig:domains}a).

\begin{figure}[t!]
    \centering
    \includegraphics[width=\columnwidth]{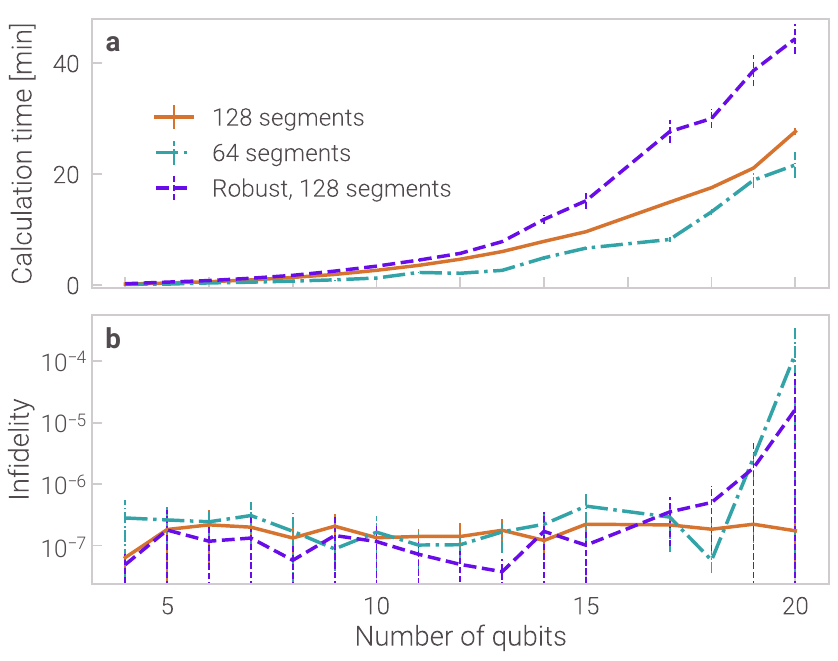}
    \caption{
    Scaling of calculation time and infidelity with ion chain length for two simultaneous maximally-entangling gates on ions (0,1) and (2,3) in the chain, using the AM$+$PM control scheme.
    Individual drives were applied to each ion for an operation duration of 300~$\mu$s, with $\Omega_\text{max}/2\pi = 100$~kHz.  A single optimization involves selecting the best performing solution from five realizations of the computation.  We then repeat this process ten times and average these results for each data point, while error bars give the standard deviation.  The data points for 16 ions in the chain have been omitted as the chosen trap frequencies give a geometric anomaly for the calculation time and fidelity with this ion number (see the SI).
    Calculations were performed on a MacBook Pro (2019) using CPU (Processor: 1.4~GHz Intel Core i5; Memory: 8~GB 2133~MHz LPDDR3).
    \label{fig:scaling}}
\end{figure}

Finally, we explore the performance-scaling of the optimization framework we employ with the number of qubits, considering both time-to-solution and minimum achievable infidelity. In this scaling analysis, we perform optimizations using chains up to 20 ions in length given state-of-the-art experimental capabilities,\cite{Innsbruck_20ions} and execute code using like-for-like local-instance hardware (a standard consumer grade laptop).  These two metrics are presented in Figure~\ref{fig:scaling} for the optimization of two parallel pairwise gates implemented within ion chains of different lengths. 

We observe that a single gate-optimization calculation may be completed via local-instance code execution in $\lesssim 50$ min for the longest 20-ion chain considered here.  Parallelization using cloud-compute infrastructure has been shown to reduce the absolute time-to-solution by a variable factor depending on the structure of the optimization, but reported up to $\sim10\times$ in previous tests~\cite{Ball2020} when leveraging GPU support for complex tasks.  Calculations require $\lesssim 10$ min in total runtime up to $\sim 14$ ions for the gate configurations treated here.  As expected, the addition of symmetry constraints in robust optimizations adds only a small overhead for chains $\leq 12$ ions in length, with an approximate doubling of runtime for longer chains.  We find that within the range of parameters considered runtimes also scale approximately linearly with segment number.  In all cases (except for the 19- and 20-ion chains) achieved infidelities are $\sim 10^{-7}$. For the longest chains, the availability of 64 unconstrained drive segments (128 for the robust case) over which the optimization is performed appears insufficient to obtain a baseline infidelity equivalent to that achieved for shorter ion chains.

\section{Conclusions and Outlook}

This work addressed the problem of achieving reconfigurable, high-fidelity multi-qubit gates in large trapped-ion registers. By framing the problem of obtaining target quantum gates using complex drives, and exploiting computationally efficient numeric optimization, we obtain the most flexible control solutions reported in the literature to the best of our knowledge.
Specifically, the control solutions we demonstrate employ both phase and amplitude modulation on the mediating laser field implementing a M\o lmer-S\o rensen interaction.  Numerically optimized solutions enact high-fidelity multi-body and parallel operations by development of a cost function which includes both motional decoupling and achievement of target pairwise entangling phases.  We have realized solutions on chains of up to 20 ions in this work, demonstrated the ability to engineer robustness to common sources of laser noise, and incorporated common constraints on modulation hardware into the optimization procedure.  These highly-configurable operations exhibit faster gate times (or lower power requirements) than controls with only amplitude modulation, and time-to-solution remains manageable for standard computational resources available in consumer laptops.

Implementation of quantum logic gates in large trapped-ion registers requires that the gate constructions be fast, flexible, high-fidelity, and scalable in order to leverage the benefits of trapped-ion hardware.  This work has contributed to each of these desiderata, while maintaining a focus on addressing practical implementation challenges.  We are excited to extend this framework to incorporate new forms of error robustness including nonlinearities in modulator response, laser-amplitude fluctuations, and laser crosstalk.  The software-configurable nature of interactions in trapped-ion quantum computers~\cite{Linke3305} makes them an ideal target for advanced numeric optimization techniques and we look forward to continuing to advance the utility of quantum optimal control techniques in this hardware platform.

\section*{Acknowledgments}
The authors are grateful to experimental colleagues at the University of Sydney:  C. Edmunds, C. Hempel and A. Milne for guidance on trap parameters.  

\section*{Conflict of Interest}
The authors declare no conflict of interest.

\begingroup
\renewcommand{\section}[2]{}
\bibliographystyle{unsrtnat-AQT}
\bibliography{bibliography}
\endgroup

\clearpage
\onecolumngrid

\section*{Table of Contents}
Numeric optimization and control techniques are demonstrated to create a wide range of fast, robust, high-fidelity gates. Complex drive controls, with both phase and amplitude modulation on the mediating laser field, are applied to enact multi-body and parallel operations on chains of up to 20 ions. Control solutions incorporate real constraints on modulation hardware and robustness to laser noise sources. \\

ToC keyword: quantum control

\begin{figure}[ht!]
    \centering
    \includegraphics{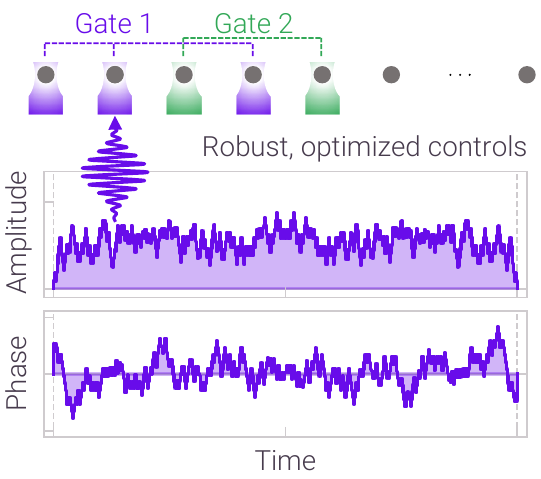}
\end{figure}

\clearpage
\appendix

\section*{Supporting Information}
\renewcommand{\thesubsection}{\Alph{subsection}}

\subsection{Fidelity derivation} \label{app:fidelity}

Here we derive the operational fidelity, which is given by
\begin{equation}
    \mathcal{F}_{av} = \left| \frac{1}{D} \text{Tr} [E] \right|^2,
\end{equation}
with $E = U_c^\dagger U (\tau)$ and where $E$ has dimension $D$. 

Observing that $\sigma_x$ operators commute, we separate $U(\tau)$ from equation~(\ref{eq:ions_unitary}) into two parts $U = U_1 U_2$:
\begin{align}
    U_1 (\tau) &= \exp \{i \sum_{j=1}^N \sum_{k=1}^{j-1} (\phi_{jk} (\tau)+\phi_{kj} (\tau)) \sigma_{x,j} \sigma_{x,k} \}\\
    U_2 (\tau) &= \exp \{ \sum_{j=1}^N \sigma_{x,j} B_j (\tau) \} \\
    &= \prod_{p=1}^M \mathcal{D}_p \left( \sum_{j=1}^N \sigma_{x,j} \eta_j^p \alpha_j^p (\tau) \right),
\end{align}
with displacement superoperator $\mathcal{D}_p (\kappa) = \exp \{ \kappa a_p^\dagger - \kappa^* a_p \}$.

Then
\begin{align}
    U_c^\dagger U_1 = \prod_{\substack{j=1 \\ k<j}}^N \left[ \cos (\epsilon_{jk}) - i \sin (\epsilon_{jk}) \sigma_{x,j} \sigma_{x,k} \right], \label{eq:ions_fid1}
\end{align}
where $\epsilon_{jk} \equiv \psi_{jk} - (\phi_{jk} (\tau) + \phi_{kj} (\tau))$.
We now take the expectation value of the ion motion with respect to a separable thermal product state:
\begin{align}
    \rho_m &= \rho_m^1 \otimes ... \otimes \rho_m^M, \\
    \rho_m^p &= \left( 1-e^{-\hbar \nu_p/kT} \right) \sum_{n=0}^\infty |n \rangle \langle n| e^{-n \hbar \nu_p/kT},
\end{align}
noting from~[\onlinecite{RoosNJP2008}] that:
\begin{equation}
    \langle \mathcal{D}_p (z_p) \rangle_{\rho_m^p} = e^{-|z_p|^2 \left(\overline{n}_p+\frac{1}{2} \right)},
\end{equation}
where $\overline{n}_p$ is the mean phonon occupation in mode $p$.
Thus we have
\begin{align}
    \langle U_2 (\tau) \rangle_{\rho_m} = \prod_{p=1}^M e^{- \left| \sum_{j=1}^N \sigma_{x,j} \eta_j^p \alpha_j^p (\tau) \right|^2 \left(\overline{n}_p+\frac{1}{2} \right)}. \label{eq:ions_fid2}
\end{align}

Returning to $E = U_c^\dagger U_1 U_2$, and taking the expectation value with respect to the motional state, we can combine the components in equations~(\ref{eq:ions_fid1}) and (\ref{eq:ions_fid2}).
Keeping diagonal terms in the internal state up to second order in the error terms $\epsilon_{jk}$ and $\alpha_j^p (\tau)$, we obtain the fidelity equation~(\ref{eq:fidelity}) in the main text.

\subsection{Asymmetric laser-detuning induced dephasing} \label{app:dephasing}

To analyse errors in the laser detunings such that $\delta_a \neq -\delta_b$ for lasers $a$ and $b$, we need to modify the derivation of equation~(\ref{eq:ions_greenform}). 
Following the Lamb-Dicke approximation and rotating wave approximations (for $\delta_a \sim \nu_p$) we obtain the interaction Hamiltonian:
\begin{align}
    H_I = \frac{\hbar \Omega}{2} \sum_{j=1}^N \left( i \sigma_{+,j} \sum_{p=1}^M \left[ \eta_j^p \left( a_p e^{-i((\nu_p + \delta_b)t - \phi_b)} + a_p^\dagger e^{i((\nu_p - \delta_a)t + \phi_a)} \right) \right] + \text{H.c.} \right),
\end{align}
for laser $a$ ($b$) with detuning $\delta_{a(b)}$ and phase $\phi_{a(b)}$.
Introducing errors $\epsilon_\delta$ and $\epsilon_\phi$ such that $\delta_b = -\delta_a + \epsilon_\delta$ and $\phi_b = -\phi_a - \pi + \epsilon_\phi$, and with $\delta_p = \nu_p - \delta_a$, we obtain
\begin{align}
    H_I = \frac{\hbar \Omega}{2} \sum_{j=1}^N \bar{\sigma}_{x,j}(t) \sum_{p=1}^M \left[ -\bar{\beta}_j^{p*} a_p + \bar{\beta}_j^p a_p^\dagger \right],
\end{align}
where
\begin{align}
    \bar{\sigma}_{x,j}(t) &= \sigma_{+,j} e^{-i \epsilon_\delta t/2 + i \epsilon_\phi/2} + \sigma_{-,j} e^{i \epsilon_\delta t/2 - i \epsilon_\phi/2}, \\
    \bar{\beta}_j^p &= \eta_j^p \frac{\Omega}{2} e^{i \phi_a} e^{i \delta_p t} e^{i \epsilon_\delta t/2 - i\epsilon_\phi /2} = \beta_j^p e^{i \epsilon_\delta t/2 - i\epsilon_\phi /2}.
\end{align}

The revised unitary evolution becomes
\begin{align}
    U (\tau) &= \mathcal{T} \exp \left(
    \sum_{j=1}^N \sum_{p=1}^M \left( \bar{\alpha}_j^p (\tau) a_p^\dagger - \bar{\alpha}_j^{p*} (\tau) a_p \right) + i \sum_{j=1}^N \sum_{k=1}^{j-1} \bar{\phi}_{jk} (\tau) \right), \label{eq:ions_dephasing_U} \\
    \bar{\alpha}_j^p (\tau) &\equiv \int_0^\tau dt \bar{\sigma}_{x,j} (t) \bar{\beta}_j^p (t), \label{eq:ions_dephasing_motion}\\
    \bar{\phi}_{jk}(\tau) &\equiv \text{Im} \left[ \sum_{p=1}^M \int_0^\tau dt_1 \int_0^{t_1} dt_2 \bar{\beta}_j^p (t_1) \bar{\beta}_k^{p*} (t_2) \bar{\sigma}_{x,j} (t_1) \bar{\sigma}^*_{x,k} (t_2) \right]. \label{eq:ions_dephasing_phase}
\end{align}

The error contributions from imperfect phase and detuning relationships between the bichromatic lasers have been separated into a change in $\beta_j^p$ and in $\sigma_{x,j}$.
The change in $\beta_j^p$ can be seen as a systematic error in detuning $\delta_p$ and the phase $\phi_a$.
This phase error term does not impact the motion or phase conditions, while robustness to quasi-static errors in $\delta_p$ is implemented in the main text (section~\ref{sec:robust}).

For small errors $\epsilon_\delta \tau \ll \pi$ and $\epsilon_\phi \ll \pi$, we can consider the first-order expansion of the exponential terms in $\bar{\sigma}_{x,j}(t)$:
\begin{equation}
    \bar{\sigma}_{x,j}(t) \simeq \sigma_{x,j} + \sigma_{y,j} (\frac{\epsilon_\delta}{2} t - \frac{\epsilon_\phi}{2}).
\end{equation}

First consider the detuning error $\epsilon_\delta$. 
Note that the first term in the exponent of equation~(\ref{eq:ions_dephasing_U}) corresponds to the motional displacement terms in the absence of error, which are set to 0 at the end of the operation.
This is achieved including the error terms by setting
\begin{align}
    0 &= \bar{\alpha}_j^p(\tau) \quad \forall p,j \\
    &\simeq \sigma_{x,j} \int_0^\tau \bar{\beta}_j^p (t) dt + \sigma_{y,j} \frac{\epsilon_\delta}{2} \int_0^\tau t \bar{\beta}_j^p (t) dt,
\end{align}
where up to systematic error in the detuning, the first term is the usual residual motion condition, and the second term is equivalent to equation~(\ref{eq:ions_dephasing_integral}).
This equation gives rise to the center-of-mass robustness condition derived for mode frequency dephasing.

The error $\epsilon_\phi$ such that $\phi_b = -\phi_a -\pi+\epsilon_\phi$ produces first-order error terms in equation~(\ref{eq:ions_dephasing_phase}) that cannot be avoided: these $\sigma_{x,j} \sigma_{y,k}$ terms have a shared coefficient with the $\sigma_{x,j} \sigma_{x,k}$ terms that determines the acquired relative phase.
However, the motional contribution of these error terms in equation~(\ref{eq:ions_dephasing_motion}) benefits from the usual residual motion condition since the error contribution only produces coefficients for the usual mode closure condition.

We thus expect the same symmetry and center-of-mass condition approach to produce robustness to both mode frequency and laser detuning dephasing.
Note, however, that error terms remain in this laser detuning case in the relative phase terms and cross-terms from the time ordering.

\subsection{Trap and laser parameters used in calculations}

Here we provide simulation details for the results in the main text.
In this work we consider $^{171}$Yb$^+$ ions, however our methodology can be applied to different ion species.
In Figures~\ref{fig:robust} and~\ref{fig:domains},
the center-of-mass trap frequencies for the [$x$, $y$, $z$] axes are set to [1.6, 1.5, 0.3]~MHz, where $z$ is the trapping axis. 
These are revised for larger numbers of ions to maintain the linear ion arrangement: [1.6, 1.5, 0.1]~MHz in Figure~\ref{fig:largearray} and [2.0, 2.0, 0.2]~MHz in Figure~\ref{fig:scaling}.
The laser wavevector is $\mathbf{k} = (2 \pi)/(355 \times 10^{-9}) \times [1, 1, 0] \text{ rad.m}^{-1}$,
and unless otherwise stated, the laser detuning is set to 4.7~kHz above the $x$-axis center-of-mass frequency.

\subsection{Scaling analysis and geometry anomaly} \label{app:16ions}

\begin{figure}[ht!]
    \centering
    \includegraphics{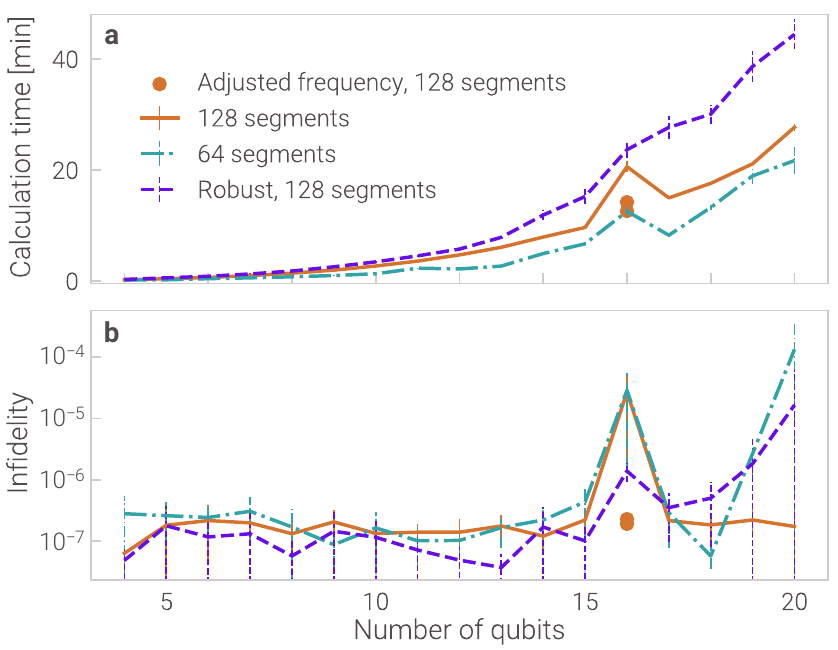}
    \caption{
    Scaling of calculation time and infidelity with ion chain length for two simultaneous maximally-entangling gates on ions (0,1) and (2,3) in the chain, using the AM$+$PM control scheme.
    Parameters match those in Figure~\ref{fig:scaling} in the main text.
    The added 'Adjusted frequency' points have the axial center-of-mass frequency set to 0.19 and 0.21~MHz (adjusted from 0.2~MHz shown on the main curve), and are the average values from ten optimizations.
    \label{fig:scaling_with16}}
\end{figure}

In the main text, we noted that the 16-ion point was removed from Figure~\ref{fig:scaling} as it is anomalous for trap geometry reasons.
Figure~\ref{fig:scaling_with16} contains the data from Figure~\ref{fig:scaling} along with additional detail.
We include the anomalies in calculation time and fidelity; for this length of chain it is particularly difficult (in calculation time and fidelity) to find effective control solutions.
We also include control performance where we vary the axial mode frequency by a small fraction for the 128-segment scheme, which then provides more expected calculation times and infidelities.
These results demonstrate that the 16-ion anomaly is not fundamental to 16-ion chains but rather a quirk of the given trap values.
Note also that these results do not impact the discussion in the main text regarding the typical scaling of required resources for finding high-fidelity control solutions.

\end{document}